\documentclass[aps,prl,footinbib,preprintnumbers,a4paper,tightenlines,superscriptaddress, floatfix,onecolumn,11pt]{revtex4}
\usepackage[colorlinks,linkcolor=blue, urlcolor=blue,anchorcolor=blue,citecolor=blue]{hyperref}
\usepackage{graphicx}
\usepackage{longtable}
\usepackage{amsmath}
\usepackage{amssymb}
\usepackage{subfigure}
\usepackage{cleveref}
\usepackage{dcolumn}
\usepackage{bm}
\usepackage{color}
\usepackage{fancyhdr}
\usepackage{epstopdf}
\allowdisplaybreaks
\usepackage{enumitem}

\begin{document}
\title{A Three-Dimensional Laser Interferometer Gravitational-Wave Detector}
\author{Mengxu Liu}
\affiliation{Physics Department, Huazhong University of Science and Technology, Wuhan 430074, China}
\author{Biping Gong}
\email[Email: ]{bpgong@hust.edu.cn}
\affiliation{Physics Department, Huazhong University of Science and Technology, Wuhan 430074, China}

\begin{abstract}

\section{Abstract}
The gravitational wave (GW) has opened a new window to the universe beyond the electromagnetic
spectrum. Since 2015, dozens of GW events have been caught by the ground-based GW detectors
through laser interferometry. However, all the ground-based detectors are L-shaped Michelson
interferometers, with very limited directional response to GW. Here we propose a three-dimensional
(3-D) laser interferometer detector in the shape of a regular triangular pyramid, which has more
spherically symmetric antenna pattern. Moreover, the new configuration corresponds to much
stronger constraints on parameters of GW sources, and is capable of constructing null-streams to get
rid of the signal-like noise events. A 3-D detector of kilometer scale of such kind would shed new light
on the joint search of GW and electromagnetic emission.

\end{abstract}

\maketitle

\section{Introduction}

Gravitational waves produced by dynamic acceleration of celestial objects are direct predictions of Einstein's General Theory of Relativity.
Together with
the electromagnetic radiation,  the physics of GW events can be investigated in a   depth that has never achieved before. In the past decades, many GW detectors have been proposed and constructed, including the  ground-based and the space-based detectors for various wavelength of gravitational waves \cite{2014RvMP...86..121A}. The basic idea of those gravitational wave detectors is to measure the relative displacement of the freely falling bodies through laser interferometetry.
Currently, most GW detection on the ground
are performed in the high frequency band (10Hz-100kHz),  by the long arm laser
interferometers, such as  TAMA 300 m interferometer \cite{PhysRevLett.86.3950}, the GEO 600 m
interferometer \cite{2006CQGra..23S..71L}, and the kilometer size laser-interferometric GW detectors like Advance LIGO (4 km arm length) \cite{2015CQGra..32g4001L}, Advance VIRGO (3 km arm length) \cite{Acernese_2014}, and the following ET (10 km arm length) \cite{2010et}.

\begin{figure}[htb]
\includegraphics[width=0.5\textwidth]{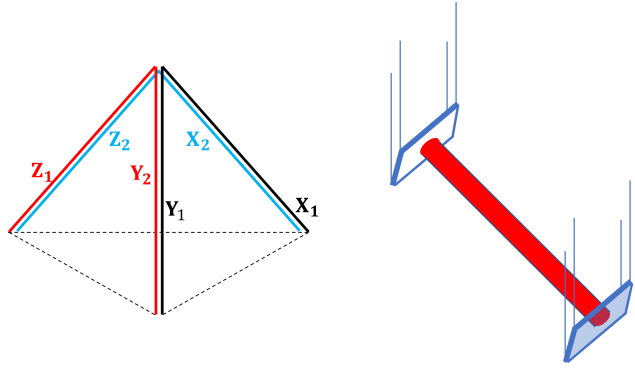}
\caption{Sketchy configuration of the 3-D interferometer. Left panel: geometry of the three sub-interferometers  in a shape of regular triangular pyramid. Right panel: configuration of one arm. }\label{fig:3D}
\end{figure}

Up to now, interferometer detectors on the ground are all  L-shaped, and most of them are Michelson interferometers. This kind of detectors has quite limited sensitivity for specific directions, namely, blind directions, which is the main reason why the Advance VIRGO failed to detect the GW170817 event \cite{2017PhRvL.119p1101A}. Besides, such detectors provide quite limited position information, so that GW sources are located through network of detectors. The coming third generation detector ET, with a geometry of triangular shape in a 2-D plane, can improve these situations \cite{2009Freise}.  Here,  a 3-D laser interferometer, with a regular triangular pyramid as shown in \cref{fig:3D},  is proposed. The additional dimension endows it with merits that the traditional L-shaped interferometers and the ET detectors don't have.

\section{Structure}
\begin{figure}
\includegraphics[width=0.6\textwidth]{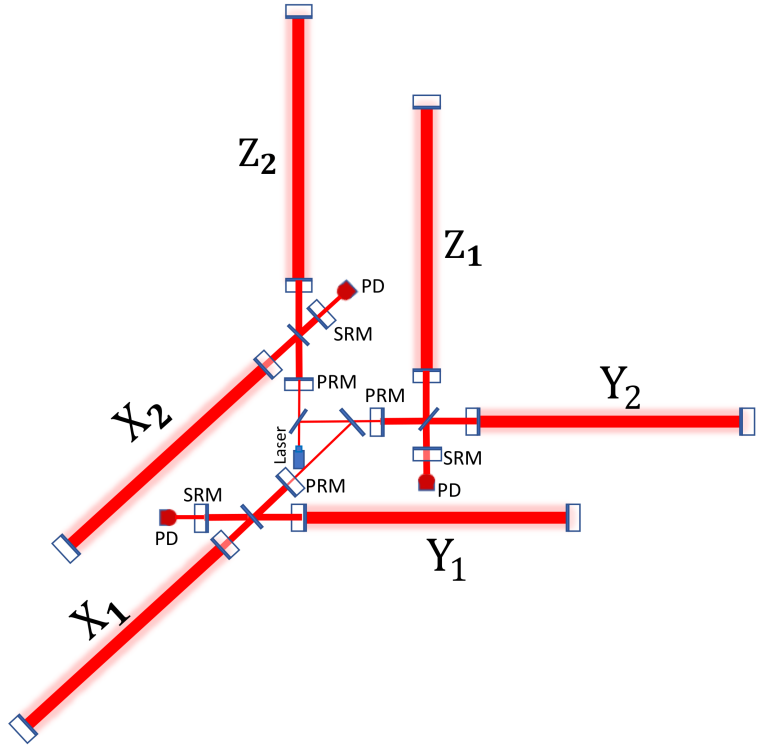}
\caption{Optical configuration of the 3-D interferometer. \{$X,Y,Z$\} indicates the arm cavities. Note the arms are not in the same plane.  PRM: power recycling mirror; SRM: signal recycling mirror; PD: photodetector.}\label{fig:opt_config}
\end{figure}

The set up of the 3-D detector as shown in \cref{fig:3D} can be reploted as  three Michelson interferometers mounting on  three planes perpendicular to each other, so that each  axis lies two arms as shown in \cref{fig:opt_config}.
 A Fabry-Perot resonant cavity in each arm is used to enhance the phase shift produced by  the change of arm length originating in gravitational radiation. The power recycling system and the signal recycling system are also involved in order to strengthen the light power inside the interferometer and widen the arm cavity bandwidth for the signal sidebands.

Three interferometers, $X_1-Y_1$, $Y_2-Z_1$, and $X_2-Z_2$  can work independently. By sharing the common laser source, three interferometers possess the same laser stability.
Besides, such a configuration guarantee that the gravity effect on the two arms of each interferometer is approximately equivalent. Moreover, all the mirrors in the detector can be mounted in an ultra-high vacuum system on the seismically isolated platforms. As a result, the same noise environment in each arm is maintained at the utmost.

\section{pattern functions} Such a  3-D laser interferometer can  greatly improve  the detector's response to GW strain. As an interferometer has different sensitivity to GWs from different directions, a 3-D detector owns specific pattern functions deviating  a L-shaped one.

The response of an interferometer to GWs is the difference between relative length changes of the two arms which can be computed from the formula
$h=\frac{1}{2}(\boldsymbol n_1 \cdot [ H \boldsymbol n_1]- \boldsymbol n_2 \cdot [ H \boldsymbol n_2])$\cite{pattern_function}, where $\boldsymbol n_1$ and $\boldsymbol n_2$ denote the unit vectors of arm direction. The three arms of the 3-D interferometer are along the $x, y $ and $z$ axes defining  the coordinate system of the detector. Another reference frame $(x', y', z')$ represents the GW's coordinate, in which the $z'$ axis stands for the propagation direction of GW. Then in the frame $(x', y', z')$ GW has the form,
\begin{equation}
H'=\left(
 \begin{array}{ccc}
  h_+(t) & h_{\times}(t) & 0\\
  h_{\times}(t) & -h_+ (t) & 0\\
  0 & 0 & 0\\
  \end{array}
  \right).
\end{equation}

The frame $(x',y',z')$ can be achieved through the rotation of the frame $(x,y,z)$, with a rotation matrix of

\begin{equation}
R=\left(
 \begin{array}{ccc}
  cos \phi & -sin \phi & 0\\
  sin \phi & cos \phi  & 0\\
  0 & 0 & 1\\
  \end{array}
  \right)
  \left(
  \begin{array}{ccc}
  1 & 0 & 0\\
  0 & cos \theta  & -sin \theta\\
  0 & sin \theta & cos \theta\\
  \end{array}
  \right)
  \left(
  \begin{array}{ccc}
  cos \psi & -sin \psi & 0\\
  sin \psi & cos \psi  & 0\\
  0 & 0 & 1\\
  \end{array}
  \right),
\end{equation}
where $\theta$ and $\phi$ denote the polar and azimuthal coordinates of the source relative to the antenna respectively, and $\psi$ represents  the polarization angle of the GW.

Then GW in the frame $(x, y, z)$ is given by $H_{ij}=R_{ik} R_{jl} H_{kl}^{'}$. As a result, the response function $h$ can be expressed in the form
\begin{equation}
h=F_+ h_+(t) + F_{\times} h_{\times}(t).
\end{equation}

\begin{figure}
\includegraphics[width=0.7\textwidth]{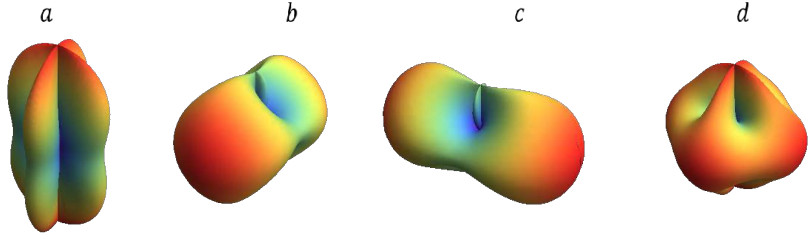}
\caption{Assuming polarization angle $\psi=0$,the response of the interferometer to (+) polarization on three planes are shown respectively, $X$-$Y$ plane [$a$],  $Y$-$Z$ plane [$b$], $X$-$Z$ plane [$c$], and the total response to (+) polarization [$d$]. Color indicates increasing sensitivity from indigo to red.}\label{fig:h_plus}
\end{figure}

The pattern functions are obtained:
\begin{align}
{\rm for} \ &  X \  {\rm and} \ Y \ {\rm arms},\nonumber\\
      &F_+(\theta,\phi,\psi)=\frac{1}{2}(1+cos^2 \theta)\ cos 2\phi\,cos 2\psi -cos \theta\,sin 2\phi\,sin 2\psi,\label{eq:5}\\
      &F_{\times}(\theta,\phi,\psi)=-\, \frac{1}{2}\,(1+cos^2 \theta) cos 2\phi\,sin 2\psi\,- sin 2\phi\,cos \theta\,cos 2\psi.\label{eq:6}\\
{\rm for} \ &  Y \  {\rm and} \ Z \ {\rm arms},\nonumber\\
      &F_+(\theta,\phi,\psi)=\frac{1}{2}((sin^2 \phi + sin^2 \theta -cos^2 \phi\,cos^2 \theta )cos 2\psi +cos \theta\,sin 2\phi\,sin 2\psi),\label{eq:7}\\
      &F_{\times}(\theta,\phi,\psi)=\frac{1}{2} ((-sin^2 \phi - sin^2 \theta +cos^2 \phi\  cos^2 \theta )sin 2\psi + sin2\phi\,cos\theta\,cos 2\psi)\label{eq:8}.\\
{\rm for} \ &  X \  {\rm and} \ Z \ {\rm arms},\nonumber\\
      &F_+(\theta,\phi,\psi)=\frac{1}{2}((cos^2 \phi + sin^2 \theta -sin^2 \phi\,cos^2 \theta )cos 2\psi -cos \theta\,sin 2\phi\,sin 2\psi),\label{eq:9}\\
      &F_{\times}(\theta,\phi,\psi)=\frac{1}{2} ((-cos^2 \phi - sin^2 \theta +sin^2 \phi\  cos^2 \theta )sin 2\psi - sin2\phi\,cos\theta\,cos 2\psi)\label{eq:10}.
\end{align}
 Assuming $\psi=0$, the pattern functions for the (+) polarization are illustrated  by a spherical polar plot, as is shown in  \cref{fig:h_plus}. The response of a traditional L-shaped detector to the (+) polarization is as shown in panel a of  \cref{fig:h_plus}. In contrast, the total response of the 3-D detector to the (+) polarization is as shown in panel d of  \cref{fig:h_plus}, apparently the response to the (+) polarized GW in directions perpendicular to Z axis is strengthened.  Consequently, the 3-D detector can provide a more isotropic antenna pattern than the conventional L-shaped detector (as shown in the panel a of \cref{fig:h_plus}), which corresponds to wider field of view to GW sources comparing to previous detectors.

\section{Benefits}

An important benefit that the 3-D detector can bring about  is the construction of null-streams \cite{null-stream}. A troublesome problem encountered GW detection is how to distinguish the signals from the noise events. A simple but powerful method called construction of null-streams can  solve such a problem, in which all the data are linearly combined in order to eliminate the GW signals. Then the output of null-streams will only be noise. Comparing with the pattern functions, one can easily find the relation
\begin{equation}
h_{XZ}=h_{XY}+h_{YZ},
\end{equation}
which shows that the 3-D detector is redundant and null-streams can be generated by the linear combination of the output from three sub-interferometers.

Another attractive benefit of such a 3-D detector is more stringent  constraints on parameters of GW sources comparing those given by previous detecors. As we know, an individual GW signal (non-spinning) depends on nine parameters: two masses $\mathcal{M}_z$ and $\mu_z$, position angles $\theta$ and $\phi$, orientation angles of the binary $\iota$ and $\psi$, time at coalescence $t_c$, phase at coalescence $\Phi_c$, and luminosity distance $D_L$. Measuring the GW phase can determine $\{\mathcal{M}_z,\mu_z, t_c,\Phi_c\}$, while the remaining five parameters cannot be individually constrained \cite{2010ApJ...725..496N}.  A single Michelson interferometer can constrain the five parameters by two quantities, while the 3-D detector constrains the five parameters by  four quantities. This markedly reduces the parameter space of a GW event.  

\begin{figure}
\includegraphics[width=0.75\textwidth]{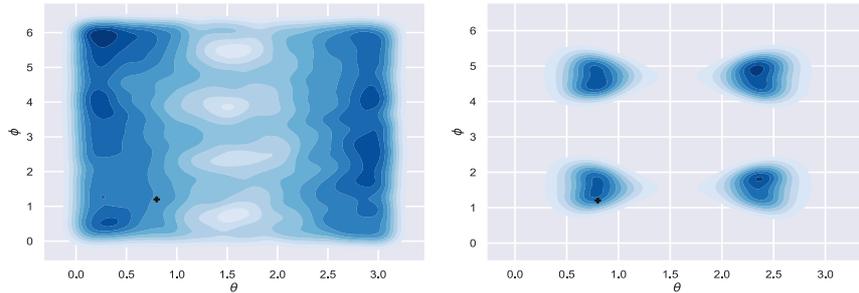}
\caption{Simulated probability density distributions for position angles $ \{\theta,\phi\}$. Left panel: the traditional L-shaped interferometer. Right panel: the 3-D interferometer. The true values ($\theta=0.8,\phi=1.2$) are marked with a cross, and the outer contour shows the $95\%$ confidence range. }\label{fig:PDF}
\end{figure}

 The reduction of the positional uncertainty by the 3-D interferometer can be simulated. We choose a set of values for the five parameters and assume that an observation can offer the probability distributions of the two quantities,
$F_+ (1+cos^2 \iota)/D_L$ and $F_{\times} cos\iota/D_L$.
 With Markov Chain Monte Carlo (MCMC) techniques we can constrain the distributions of position angles $\{\theta,\phi\}$  as shown in \cref{fig:PDF}. Due to the extra pattern functions presented in the 3-D interferometer, the positional uncertainty can be significantly reduced comparing with the traditional L-shaped interferometers. In particular,  the position of the GW source is restricted into four regions by the 3-D detector so that the position angles will become much more explicit. While a complete reconstruction of the parameters of GW sources will need  a joint observation of another detector at a different location.

\begin{figure}
\includegraphics[width=0.45\textwidth]{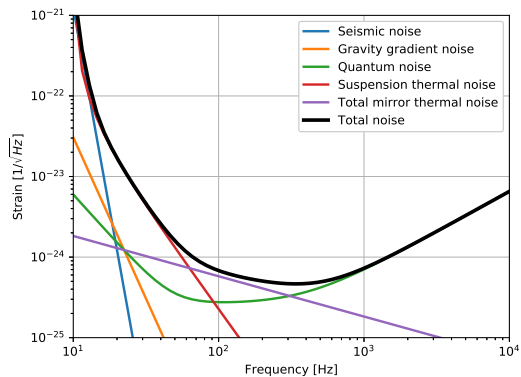}
\caption{Estimated sensitivity of the 3-D interferometer. }\label{fig:strain}
\end{figure}

\section{Sensitivity estimation} The level of total noise  determines the weakest GW signals  detectable. As a matter of fact, the detector will suffer several fundamental noises, including quantum noise \cite{quantum}, thermal noise \cite{thermal_noise}, seismic noise \cite{seismic}, and gravity gradient noise \cite{gravity_gradients} et al. Here we estimate the sensitivity limit of the 3-D detector based on the idealized  parameters which are likely to come true in the future.

 Referring to other laser interferometer detectors, the parameters of the 3-D detector are assumed as follows. The length of each arm is set as 8 km, the mass of each mirror 200 kg, the loss angle of the coating $ 5\times 10^{-5}$, the loss angle of the substrate $5 \times 10^{-9}$. And the whole system should be operated at temperature 290 K.
Moreover,  each mirror is suspended by the quadruple pendulum with the resonant frequency of 10 Hz and loss angle of $10^{-9}$.
Besides, a Fabry-Perot cavity with a fineness of 1000 is placed in each arm, and the laser power in the cavity  will  be 2 MW.
In this case, the corresponding noise estimation can be calculated, as shown by the black curve  in \cref{fig:strain}. In such a case, the sensitivity of the 3-D detector  is improved roughly by a factor of 10 in contrast with the designed sensitivity of Advanced LIGO in high frequency range \cite{2015CQGra..32g4001L}. The strain noise curve in \cref{fig:strain} corresponds to the binary neutron star (BNS) range around 1000 Mpc.

\section{Discussion}

The high sensitivity of such a detector can support broader scientific goals relative to the second generation detectors. The higher signal-to-noise ratio (SNR) and higher probability of  identification of GW events  will be useful in answering  questions, like the origin and evolution of black hole and the inner structure of NS. The detector  may even detect the signals from new astrophysical sources like core collapse supernovae and isolated rotating NSs. Because of its high sensitivity in high frequency band, the detection of post-merger signals spanning from 1 to several kHz from the BNSs is expected. Undoubtedly, the detection of the peak frequencies of the post-merger stage can play a crucial role in constraining the equation of state of neutron stars \cite{Chatziioannou_2017,Bose_2018,Torres_Rivas_2019}. While, the behavior of such a detector at low frequency is not as good as at high frequency due to the  thermal noise. To suppress the low frequency noise, one may consider the xylophone configuration by adding another instrument operated with low power and cryogenic mirrors\cite{10xylophone}.

As for  the implementation,  the construction of a kilometer-scale 3-D detector and keeping the stability of such a detector maybe a tough problem to handle. Here we put forward a plausible way  to build the detector by taking advantage of the mountain's terrain, then two arms in each direction will be packaged by a tunnel, and the tunnel can be extended to under-ground to accommodate long arms needed by this design. However, this scheme may have some issues related with seismic noise and gravity noise which mainly dominate the low frequency range. The geometry of the  3-D detector is far different from the traditional L-shaped detector. Therefore, a 10-meter prototype  may be useful to test the new way of suspension and the alignment control system.

Although the  triangular shaped ET can also provide the null-streams and strong constraints on the parameters of the GW sources, the 3-D detector can have a more isotropic antenna pattern excepting for few blind spots compared with ET. Due to the short duration of the merged signals (up to several minutes), it is not realistic to only count on the earth rotation to widen the field of view. Therefore, a more isotropic antenna pattern is  important in the increase of  detection rate.
 In addition, the opening angle of the sub-interferometers in  ET is $60^{\circ}$, while for the 3-D is $90^{\circ}$, which resulting in the $\frac{SNR_{ET}}{SNR_{3D}}=sin(60^{\circ})\approx 0.87$ if assuming the same arm length. So the 3-D detector can achieve more isotropic antenna pattern without losing sensitivity.

\section{Conclusion}
 We propose a 3-D GW detector with three Michelson interferometers setting in a  regular triangular pyramid. The original motivation of the geometry is that such a detector would have a more spherically symmetric antenna pattern. In this paper, we have shown that the detector is fully redundant and able to generate null-streams by data from three sub-interferometers. And the detector can also provide much stronger constraints on the parameters of GW sources. So a network of the 3-D detectors will make the detection and parameter estimation more efficient and accurate. As a consequence, our understanding of fundamental  physics will be enhanced with the combination of  the electromagnetic signals and accurate GWs parameters at the same time.


\end{document}